\documentclass[runningheads]{llncs}

\usepackage{amscd} \usepackage{amssymb} \usepackage{amsfonts} \usepackage{array}

\usepackage[pdftex]{graphicx}

\usepackage[all]{xy}    
\newtheorem{assumption}{Assumption} 

\newcommand{\mi}[1]{{\mathit{#1}}}
\newcommand{\msf}[1]{{\mathsf{#1}}}
\newcommand{\mbf}[1]{{\mathbf{#1}}}

\newcommand{\sua}{\msf{STD} \cup \msf{ACUN}}

\newcommand{\std}{\mathsf{STD}}
\newcommand{\pk}{\mi{pk}}
\newcommand{\sh}{\mi{sh}}
\newcommand{\acun}{\mathsf{ACUN}}
\newcommand{\StdOps}{\mi{StdOps}}
\newcommand{\lra}{\leftrightarrow}
\newcommand{\Lra}{\Leftrightarrow}

\newcommand{\Ra}{\Rightarrow}
\newcommand{\EnCp}{\mi{EncSubt}}

\newcommand{\Vars}{\mi{Vars}}
\newcommand{\NewVars}{\mi{NewVars}}

\newcommand{\Agent}{\mi{Agent}}

\newcommand{\xor}{\mathtt{XOR}}
\newcommand{\txor}{\texttt{XOR}}
\newcommand{\xorseq}[3]{#1\oplus#2 \oplus #3}
\newcommand{\mxor}[2]{#1\oplus #2}

\newcommand{\Rules}{\mi{Rules}}

\newcommand{\Terms}{\mi{Terms}}

\newcommand{\appl}{\mathsf{appl}}
\newcommand{\satisfiable}{\mathsf{satisfiable}}

\newcommand{\normalize}{\mi{normalize}}
\newcommand{\normal}{\msf{normal}}

\newcommand{\act}{\msf{act}}
\newcommand{\fa}{\forall}
\newcommand{\ex}{\exists}
\newcommand{\simple}{\msf{simple}}

\newcommand{\semibundle}{\msf{semi\mbox{-}bundle}}
\newcommand{\conseq}{\msf{conseq}}

\newcommand{\SubTerms}{\mi{SubTerms}}

\newcommand{\munut}{\mu\mbox{-}\msf{NUT}}
\newcommand{\munutsat}{\mu\mbox{-}\msf{NUT}\mbox{-}\mi{Satisfying}}

\newcommand{\sfs}{\msf{secureForSecrecy}}
\newcommand{\lan}{\langle}
\newcommand{\ran}{\rangle}
\newcommand{\sqss}{\sqsubset}
\newcommand{\Constants}{\mi{Constants}}

\newcommand{\tacun}{\textsf{ACUN}}
\newcommand{\tstd}{\textsf{STD}}
\newcommand{\vip}{\mi{vip}}
\newcommand{\VarIdP}{\mi{VarIdP}}

\newcommand{\pure}{\msf{pure}}
\newcommand{\tuple}[2]{\lan #1, ~#2 \ran}

\newcommand{\FreshVars}{\mi{FreshVars}}
\newcommand{\FreshCons}{\mi{FreshCons}}

\newcommand{\LTKeys}{\mi{LTKeys}}

\newcommand{\tmgu}[1]{{#1}\mbox{-}\mi{mgu}}

\newcommand{\SecVars}{\mi{SecVars}}
\newcommand{\SecConstants}{\mi{SecCons}}

\newcommand{\iik}{\mi{IIK}}

\newcommand{\nslxor}{\msf{nsl_{\oplus}}}

\newcommand{\penc}[2]{[#1]^{\to}_{#2}}

\newcommand{\isocs}{\mi{isocs}}
\newcommand{\subiso}{\sigma_{\mi{iso}}}
\newcommand{\chisocs}{\mi{chisocs}}
\newcommand{\subcomb}{\sigma_{\mi{comb}}}
\newcommand{\combcs}{\mi{combcs}}
\newcommand{\chcombcs}{\mi{chcombcs}}
\newcommand{\childseq}{\msf{childseq}}

\newcommand{\Th}{\mi{Th}}

\newcommand{\ssh}{\mathtt{SSH}}
\newcommand{\ssl}{\mathtt{SSL}}
\newcommand{\SUA}{\msf{S} \cup \msf{A}}
\newcommand{\STDUACUN}{\msf{STD} \cup \msf{ACUN}}

\newcommand{\XorTerms}{\mi{XorTerms}}

\begin{document}

\title{Protocol independence through disjoint encryption under Exclusive-OR}
\titlerunning{Protocol independence through disjoint encryption under Exclusive-OR}

\author{Sreekanth Malladi}
\authorrunning{Sreekanth Malladi}

\institute{
		Dakota State University \\	
		Madison, SD 57042, USA \\
		\email{Sreekanth.Malladi@dsu.edu}
}
  
\maketitle

\begin{center} \today \end{center}

\thispagestyle{empty}

\begin{abstract}

Multi-protocol attacks due to protocol interaction has been a notorious problem for  security. Gutman-Thayer proved that they can be prevented  by ensuring that encrypted messages are distinguishable across protocols, under a free algebra~\cite{TG00}. In this paper, we prove that a similar suggestion prevents these attacks under commonly used operators such as Exclusive-OR, that induce equational theories, breaking the free algebra assumption.

\end{abstract}

\section{Introduction}\label{s.intro}

It is quite common for users to simultaneously run multiple cryptographic protocols on their machines. For instance, a user might connect to a web site using \texttt{https} that uses the $\ssl{}$ protocol and also connect to another remote server using the $\ssh{}$ protocol. It is also quite common for a single protocol to consist of multiple sub-protocols.

A protocol might be secure when running in isolation, but not necessarily when running parallely with other protocols. In fact, Kelsey et al.~\cite{KSW97} showed that, for any given secure protocol, it is always possible to create  another protocol to break the original protocol. In an interesting practical study, Cremers analyzed 30 published protocols and reported that 23 of them were vulnerable to multi-protocol attacks~\cite{Cremers06}. Thus, they are a genuine and serious threat to protocol security.

In an outstanding work, Guttman-Thayer proved that, if encrypted messages are tagged with distinct protocol identifiers, multi-protocol attacks can be prevented~\cite{TG00}. For instance, if the notation $[t]_k$ denotes message $t$ encrypted with key $k$, then encryptions in the $\ssl{}$ protocol should resemble $[\ssl, t_1]_{k_1}$ and those in the $\ssh{}$ protocol should resemble $[\ssh, t_2]_{k_2}$. With such tagging in place, it will not be possible for an attacker to replay encryptions across protocols, since users would check and verify the tags upon receipt of messages.

However, Guttman-Thayer considered a basic protocol model where operators for constructing messages (such as encryption algorithms) do not induce equations between syntactically different messages, such as $[t]_k = [k]_t$. Most ``real-world" protocols such as $\ssl{}$ violate this assumption, and use operators that do induce equational theories, such as Exclusive-OR (\txor). It is extremely important to revisit Guttman-Thayer result under these operators, since such operators have been demonstrated to cause new attacks on protocols that are not possible under a free algebra~\cite{RS98}.

This is the problem we consider in this paper: We prove that a tagging scheme that is similar to Guttman-Thayer's prevents multi-protocol attacks under the \txor{} operator and the \tacun{} theory induced by it. Our proof strategy is general, and could be used for other equational theories such as {\sf ACU,Idempotence} and {\sf ACU,Inverse}. We give some intuitions for this in our conclusion.

\paragraph*{Organization.} In Section~\ref{s.framework}, we introduce our formal  framework including the term algebra, protocol model, constraint satisfaction, security properties and our main protocol design requirements. In Section~\ref{s.a-lemma}, we prove a lynchpin lemma that we use in   Section~\ref{s.main-result} to achieve the main result. We conclude with a discussion of future and related works.

\section{The Framework}\label{s.framework}

In this section, we formalize our framework to model and analyze protocols. 

\subsection{Term Algebra}\label{ss.term-algebra}

We will start off with the term algebra. We derive much of our concepts here from Tuengerthal's technical report~\cite{Tuengerthal-TR-2006} where he has provided an excellent and clear explanation of equational unification.

We denote the {\em term algebra} as $T(F,\Vars)$, where $\Vars$ is a set of variables, and $F$ is a set of function symbols or operators, called a {\em signature}. The terms in $T(F,\Vars)$ are called $F$-Terms. Further, 

\begin{itemize}
	\item $\Vars \subset T(F,\Vars)$;
	\item $(\fa f \in F)(\msf{arity}(f) > 0 \wedge t_1, \ldots, t_n \in T(F,\Vars)  \Ra f(t_1,\ldots,t_n) \in T(F,\Vars))$.
\end{itemize}

The set of nullary function symbols are called {\em constants}. We assume that every variable and constant have a ``type" such as $\Agent$, $\mi{Nonce}$ etc.

We define $F$ as $\StdOps \cup \{ \xor \} \cup \Constants$, where, 

\[	\StdOps = \{ \mi{sequence}, \mi{penc}, \mi{senc}, \pk, \sh \}.	\] 

$\mi{penc}$ and $\mi{senc}$ denote asymmetric and symmetric encryption operators respectively. $\pk$ and $\sh$ denote public-key and shared-key operators respectively. We assume that they will always be used  with one and two arguments respectively, that are of the type $\Agent$. 

We use some syntactic sugar in using some of these operators: 

\begin{eqnarray*}
  \mi{sequence}(t_1,\ldots,t_n)  &	= &	[t_1, \ldots, t_n], \\
						\mi{penc}(t,k)       &	= &  [t]^{\to}_k,   \\
						\mi{senc}(t,k) 			 &	= &  {[t]^{\lra}_k},   \\
				\xor(t_1,\ldots,t_n)		 &	=	&	t_1 \oplus \ldots \oplus t_n. 
\end{eqnarray*}

We will omit the superscripts $\lra$ and $\to$ for encryptions if the mode of encryption is contextually irrelevant.

We will write ``$t ~\msf{in} ~[t_1,\ldots,t_n]$" if $t \in \{t_1,\ldots,t_n\}$. 
We will write $t_i \prec_t t_j$ if $t_i, t_j ~\msf{in} ~t$, $t = [t_1,\ldots,t_n]$ and $i < j$. 

We define the subterm relation as follows: $t \sqss t'$ iff $t' = f(t_1,\ldots,t_n)$ where $f \in 	F$ and $t \sqss t''$ for some $t'' \in \{ t_1, \ldots, t_n \}$.

We will use functions $\Vars()$, $\Constants()$, and $\SubTerms()$ on a single term or sets of terms, that return the variables, constants and subterms in them respectively. For instance, if $T$ is a set of terms,

\[	\SubTerms(T) = \{ t \mid (\ex t' \in T)(t \sqsubset t') \}.	\]

We will now introduce equational theories and equational unification.

\begin{definition}{\em\bf [Identity and Equational Theory]}\label{d.equation-theory}
Given a signature $F$, and set of variables $\Vars$, a set of {\em identities} $E$ is a subset of $T(F,\Vars) \times T(F,\Vars)$. We denote an identity as $t \cong t'$ where $t$ and $t'$ belong to $T(F,\Vars)$. An {\em equational theory} (or simply a theory) $=_E$ is the least congruence relation on $T(F,\Vars)$, that is closed under substitution and contains $E$. i.e., 

\[
	 =_E  := 
	 \left\{ 
	 		R \mid 
	 			\begin{array}{c}
	 				R ~\mathrm{is ~a ~congruence~relation~on} ~T(F,\Vars), E \subseteq R, \mathrm{and} \\
	 				(\forall \sigma)(t \cong t' \in R \Ra t\sigma \cong t'\sigma \in R)
	 			\end{array}
	 \right\}					
\]

\end{definition}

For the signature of this paper, we define two theories, \tstd{} and \tacun{}.

The theory $\std$ for $\StdOps$-Terms is based on a set of identities between syntactically equal terms, except for the operator $\sh$: 

\begin{center}
			\begin{tabular}{rcl}
  			$\{ [t_1,\ldots,t_n]$ &	$\cong$  & $[t_1,\ldots,t_n]$, \\
%				 						[t]_k 	&	\cong	  & [t]_k, \\ 	
											$h(t)$  &  $\cong$  &  $h(t)$, \\
									$sig_k(t)$  &  $\cong$  &  $sig_k(t)$, \\
										$\pk(t)$  &  $\cong$  &  $\pk(t)$, \\
				 						$[t]_k$ 	&	 $\cong$	&  $[t]_k$, \\ 	
							$\sh(t_1,t_2)$  &  $\cong$  & $\sh(t_2,t_1)  \}$.	
			\end{tabular}
\end{center}

The theory $\acun$ is based on identities solely with the \txor{} ($\oplus$) operator: 
	
	\[ \{ t_1 \oplus (t_2 \oplus t_3) \cong (t_1 \oplus t_2) \oplus t_3 ,  \mxor{t_1}{t_2}  \cong  \mxor{t_2}{t_1}, \mxor{t}{0}  \cong  t,  \mxor{t}{t}  \cong 0 \}.
	\]

We will now describe equational unification. 

\begin{definition}{\em\bf [Unification Problem, Unifier, Unification Algorithm]}\label{d.unification}

If $F$ is a signature and $E$ is a set of identities, then an $E$-{\em Unification Problem} over $F$ is a finite set of equations 

\[	\Gamma = \left\{ \begin{array}{c}  s_1 \stackrel{?}{=}_E t_1, \ldots, s_n \stackrel{?}{=}_E t_n \end{array} \right\} \]

between $F$-terms. A substitution $\sigma$ is called an $E$-{\em Unifier} for $\Gamma$ if $(\fa s \stackrel{?}{=}_E t \in \Gamma)(s\sigma =_E t\sigma)$. $U_E(\Gamma)$ is the set of all $E$-Unifiers of $\Gamma$. A $E$-Unification Problem is called $E$-{\em Unifiable} iff $U_E(\Gamma) \neq \{ \}$.

A {\em complete} set of $E$-Unifiers of an $E$-Unification Problem $\Gamma$ is a set $C$ of idempotent $E$-Unifiers of $\Gamma$ such that for each $\theta \in U_E(\Gamma)$ there exists $\sigma \in C$ with $\sigma \ge_E \theta$, where $\ge_E$ is a partial order on $U_E(\Gamma)$.

\end{definition}

An $E$-{\em Unification Algorithm} takes an $E$-Unification Problem $\Gamma$ and returns a finite, complete set of $E$-Unifiers.

Hence forth, we will abbreviate ``Unification Algorithm" to UA and ``Unification Problem" to UP.

Two theories $=_{E_1}$ and $=_{E_2}$ are {\em disjoint} if the signatures used in the identities $E_1$ and $E_2$ have no common operators. UAs for two disjoint theories may be combined to output the complete set of unifiers for  UPs made using operators from both the theories, using Baader \& Schulz Combination Algorithm (BSCA)~\cite{BS96}. 

BSCA first takes as input a $(E_1 \cup E_2)$-UP, say $\Gamma$, and applies some transformations on them to derive $\Gamma_{5.1}$ and $\Gamma_{5.2}$ that are sets of $E_1$-UPs and $E_2$-UPs respectively. It then combines the unifiers for $\Gamma_{5.1}$ and $\Gamma_{5.2}$ obtained using $E_1$-UA and $E_2$-UA respectively, to return the unifier(s) for $\Gamma$ (see Appendix~\ref{s.BSCA}, Def.~\ref{d.Combined-Unifier}). Further, if $\Gamma$ is $(E_1 \cup E_2)$-Unifiable, then there exist $\Gamma_{5.1}$ and $\Gamma_{5.2}$ that are $E_1$-Unifiable and $E_2$-Unifiable respectively.

We give a more formal and detailed explanation of BSCA in Appendix~\ref{s.BSCA}  using an example UP, for the interested reader.

\subsection{Protocol Model}\label{ss.prot-model}

Our protocol model is based on the strand space framework~\cite{THG98}. 

\begin{definition}{{\em\bf [Node, Strand, Protocol]}}\label{d.node-strand}
	A {\em node} is a tuple $\tuple{\pm}{t}$ denoted $\pm t$ where $t \in T(F,\Vars)$. A {\em strand} is a sequence of nodes. A {\em protocol} is a set of strands.
\end{definition}

 For instance, consider the $\msf{NSL}_{\oplus}$ protocol~\cite{CKRT03}:

\begin{center}
  \begin{tabular}{ll}
      {\bf Msg 1.} $A \to B : [N_A,A ]_{pk(B)}$ \\
      {\bf Msg 2.} $B \to A : [N_A \oplus B,N_B ]_{pk(A)}$\\
			{\bf Msg 3.} $A \to B : [N_B]_{pk(B)}$
	\end{tabular}
\end{center}

Then, $\msf{NSL}_{\oplus} = \{ \mi{role_A}, \mi{role_B} \}$, where,

\noindent
\begin{center}
\begin{tabular}{rll}
$\mi{role_A}$ & $= [ +[A,N_A]_{\pk(B)}, -[N_A \oplus B,N_B]_{\pk(A)}, +[N_B]_{\pk(B)}  ]$, and\\ 
$\mi{role_B}$ & $= [ -[A,N_A]_{\pk(B)}, +[N_A \oplus B,N_B]_{\pk(A)}, -[N_B]_{\pk(B)}  ]$.
\end{tabular}
\end{center}

A {\em semi-bundle}	$S$ for a protocol $P$ is a set of strands formed by applying substitutions to some of the variables in the strands of $P$: If $P$ is a protocol, then, $\semibundle(S,P) \Ra  (\fa s \in S)((\ex r \in P; \sigma)( s = r\sigma )).$

For instance, $S = \{ s_{a1}, s_{a2}, s_{b1}, s_{b2} \}$ below is a semi-bundle for the $\msf{NSL}_{\oplus}$ protocol with two strands per role of the protocol: 

\begin{center}
\begin{tabular}{rcl}

$s_{a1}$  &  $=$  &  $[ +[a1,n_{a1}]_{\pk(B1)}, -[n_{a1} \oplus B1,N_{B1}]_{\pk(A1)}, +[N_{B1}]_{\pk(B1)} ]$, \\
$s_{a2}$  &  $=$  &  $[ +[a2,n_{a2}]_{\pk(B2)}, -[n_{a2} \oplus B2,N_{B2}]_{\pk(A2)}, +[N_{B2}]_{\pk(B2)} ] $, \\
$s_{b1}$  &  $=$  &  $[ -[A_3,N_{A3}]_{\pk(b1)}, +[N_{A3} \oplus b1,n_{b1}]_{\pk(A3)}, -[n_{b1}]_{\pk(b1)}  ] $,  \\
$s_{b2}$  &  $=$  &  $[ -[A_4,N_{A4}]_{\pk(b2)}, +[N_{A4} \oplus b2,n_{b2}]_{\pk(A4)}, -[n_{b2}]_{\pk(b2)}  ] $.

\end{tabular}
\end{center}

({\em Note}: lower-case symbols are constants and upper-case are variables).

We will assume that every protocol has a set of variables that are considered ``fresh variables" (e.g. Nonces and Session-keys). If $P$ is a protocol, then, $\FreshVars(P)$ denotes the set of fresh variables in $P$. We will call the constants substituted to fresh variables of a protocol in its semi-bundles as ``fresh constants" and denote them as $\FreshCons(S)$. i.e., If $\semibundle(S,P)$, then,

\[
\FreshCons(S) = 
\left\{ 
\begin{array}{c}
 x \mid  
 \left(
 	\begin{array}{c}
	 	\ex r \in P; s \in S; \\
	 	\sigma; X
	\end{array}
 \right)
	\left(
		\begin{array}{c}
			(r\sigma = s) \wedge (X \in \FreshVars(P)) \wedge \\ 
			(x = X\sigma) \wedge (x \in \Constants)
		\end{array}
	\right)
\end{array}
\right\}.
\]

We assume that some fresh variables are ``secret variables" and denote them as $\SecVars(P)$. We define ``$\SecConstants()$" to return ``secret constants" that were used to instantiate secret variables of a protocol: If $\semibundle(S,P)$, then, \\

\[
\SecConstants(S) = 
\left\{ 
\begin{array}{c}
 x \mid 
 \left(
 	\begin{array}{c}
	 	\ex r \in P; s \in S; \\
	 	\sigma; X
	\end{array}
 \right)
	\left(
		\begin{array}{c}
			(r\sigma = s) \wedge (X \in \SecVars(P)) \wedge \\ 
			(x = X\sigma) \wedge (x \in \Constants)
		\end{array}
	\right)
\end{array}
\right\}.
\]

For instance, $N_A$ and $N_B$ are secret variables in the $\msf{NSL}_{\oplus}$ protocol and $n_{a1}, n_{a2}, n_{b1}, n_{b2}$ are the secret constants for its semi-bundle above.

We will lift the functions $\Vars()$, $\Constants()$, and $\SubTerms()$ to strands, protocols and semi-bundles. For instance, if $P$ is a set of strands and $r \in P$, then,

\[
\begin{array}{l}
	\SubTerms(r) = \{ t \mid (\ex t')((\tuple{\_}{t'} ~\msf{in} ~r) \wedge (t \in \SubTerms(t')))	\}, \\
	\SubTerms(P) = \{ t  \mid  (\ex r \in P)(t \in \SubTerms(r)) \}.
\end{array}
\]

We also define the long-term shared-keys of $P$ as $\LTKeys(P)$, where,

\noindent
\begin{eqnarray*}
	\LTKeys(P) &  =  &  \{ x  \mid  (\ex A, B)( (x = \mi{sh}(A,B)) \wedge (x \in 		\SubTerms(P)) ) \}. \\
\end{eqnarray*}

To achieve our main result, we need to make some assumptions. Most of our assumptions are reasonable, not too restrictive for protocol design and in fact, good design practices.

As noted in~\cite{CD-fmsd08}, we first need an assumption that long-term shared-keys are never sent as part of the messages in the protocol, but only used as encryption keys. Obviously, this is a safe and prudent design principle.

Without this assumption, there could be multi-protocol attacks even when Guttman-Thayer suggestion of tagging encryptions is followed. For instance, consider the following protocols:

\begin{center}
\begin{tabular}{|c|c|}
	\hline	
	$\mbf{P_1}$ 	&	   $\mbf{P_2}$ \\  \hline
	1. $a \to s : \sh(a,s)$		&		  1. $a \to b : [1,n_a]_{\sh(a,s)}$ \\
	\hline
\end{tabular}
\end{center}

Now the message in the second protocol could be decrypted and $n_a$ could be derived when it is run with the first protocol.

To formalize this assumption, we define a relation {\em interm} denoted $\Subset$ on terms such that, 

\begin{itemize}

	\item $t \Subset t'$ if $t = t'$, 
	\item $t \Subset [t_1,\ldots,t_n]$ if $(t \Subset t_1 \vee \ldots \vee t 								\Subset  t_n)$,
	\item $t \Subset [t']_k$ if $(t \Subset t')$,
	\item $t \Subset t_1 \oplus \ldots \oplus t_n$ if $(t \Subset t_1) \vee \ldots 					\vee (t \Subset t_n)$.

\end{itemize}

Notice that an interm is also a subterm, but a subterm is not necessarily an interm. For instance, $n_a$ is an interm and a subterm of $n_a \oplus [a]^{\to}_{n_b}$, while $n_b$ is a subterm, but not an interm.

Interms are useful in referring to the plain text of encryptions or everything that can be ``read" by the recipient of a term. Contrast these with the keys of encrypted terms, which can only be confirmed by decrypting with the corresponding inverses, but cannot be read (unless included in the plain-text).

\begin{assumption}\label{a.LTKeys}
	If $P$ is a protocol, then, there is no term of $P$ with a long-term key as an interm: 
	
	\[ (\fa t \in \SubTerms(P))( (\nexists t' \Subset t)(t' \in \LTKeys(P)) ). \]
	
\end{assumption}

It turns out that this assumption is not sufficient. As noted by an anonymous reviewer of this workshop, we also need another assumption that if a variable is used as a subterm of a key, then there should be no message in which that variable is sent in plain (since a long-term shared-key could be substituted to the variable as a way around the previous assumption).

Hence, we state our next assumption as follows:

\begin{assumption}\label{a.Key-Var}

	If $[t]_k$ is a subterm of a protocol, then no interm of $k$ is an interm of the protocol:
	
	\[	
		(\fa [t]_k \in \SubTerms(P))
				( (\nexists X \Subset k; t' \in \SubTerms(P))(X \Subset t')
				).
	\]

\end{assumption}

Next, we will make some assumptions on the initial intruder knowledge. We will denote the set of terms known to the intruder before protocols are run, $\iik$. We will first formalize the assumption that he knows the public-keys of all the agents:

\begin{assumption}\label{a.IIK1}
	$(\fa x \in \Constants)(\pk(x) \in \iik)$.
\end{assumption}

In addition, we will also assume that the attacker knows the values of all the constants that were substituted by honest agents for all the non-fresh variables (e.g. agent identities $a, b$ etc.), when they form semi-strands:

\begin{assumption}\label{a.IIK2}

Let $P$ be a protocol. Then, 
	
	\[	(\fa x/X \in \sigma; r \in P)
	\left(
	\left(
	\begin{array}{c}
		 \semibundle(S,P) \wedge 
		 (r\sigma \in S) \wedge \\
		 (x \in \Constants)  \wedge 
		 (X \notin \FreshVars(P))
	\end{array}
	\right)
	\Ra
					(x \in \iik)
	\right).
	\]

\end{assumption}

Finally, we make another conventional assumption about protocols, namely that honest agents do not reuse fresh values such as nonces and session-keys:

\begin{assumption}\label{a.freshness}
	
		Let $S_1, S_2$ be two different semi-bundles. Then, 
			\[	\FreshCons(S_1) \cap \FreshCons(S_2) = \{ \}.	\]

\end{assumption}

\subsection{Constraints and Satisfiability}\label{ss.constraints}

In this section, we will formalize the concepts given in \cite{MS01,Chev04} to generate symbolic constraints from node interleavings of semi-bundles and the application of reduction rules to determine satisfiability of those constraints.

\begin{definition}{{\em\bf [Constraints, Constraint sequences]}}\label{d.constraints}
A {\em constraint} is a tuple $\tuple{m}{T}$ denoted $m:T$, where $m$ is a term called the {\em target} and $T$ is a set of terms called the {\em term set}. If $S$ is a semi-bundle, then, $cs$ is a {\em constraint sequence} of $S$, or $\conseq(cs,S)$	if every target term in $cs$ is from a $-$ node of $S$ and every term in every term set in $cs$ is from a $+$ node of $S$.

\end{definition}	
	
A constraint sequence $cs$ is {\em simple} or $\simple(cs)$ if all the targets are variables. Constraint $c$ is an ``active constraint" of a constraint sequence $cs$ (denoted $\act(c,cs)$) if all its prior constraints in $cs$, but not itself,  are simple constraints. We denote the sequences before and after the active constraint of a sequence $cs$ as $cs_<$ and $cs_>$ respectively.

In Table~\ref{t.rules}, we define a set of symbolic reduction rules,  $\mi{Rules}$, that can be applied on the active constraint of a constraint sequence.

\begin{table*}
\begin{center}
\begin{tabular}{|c|c|c||c|c|c|}
	\hline 

  {\sf concat}   &   $[t_1,\ldots,t_n]:T$   &   $t_1:T$,\ldots,$t_n:T$   &   {\sf split}   &   $t:T \cup [t_1,\ldots,t_n]$	&	 $t:T \cup t_1 \cup \ldots \cup t_n$ \\  \hline
 
  {\sf penc}   &   $[m]^{\to}_k:T$   &   $k:T,m:T$   &   {\sf pdec}   &   $m:[t]^{\to}_{\mi{pk}(\epsilon)}\cup T$	  &	 $m:t\cup T$ \\  \hline

  {\sf senc}   &   $[m]^{\leftrightarrow}_k:T$   &   $k:T,m:T$   &   {\sf sdec}   &   $m:[t]^{\leftrightarrow}_k\cup T$	&	 $k:T,m:T\cup\{t,k\}$ \\  \hline

  $\msf{xor_r}$   &   $m : T \cup t_1 \oplus \ldots \oplus t_n$   &   $t_2 \oplus \ldots \oplus t_n : T,$  &   $\msf{xor_l}$   &    $t_1 \oplus \ldots \oplus t_n : T$	 & $t_2 \oplus \ldots \oplus t_n : T$,  \\  
 
 &		&		$m : T \cup t_1$ &		&		&	 $t_1 : T$ 	 \\ \hline

\end{tabular}
\end{center}

\caption{Set of reduction rules, {\em Rules}}\label{t.rules}

\end{table*}

The first column is the name of the rule, the second and third columns are the active constraints before and after the application of the rule.

We define a predicate $\appl()$ on each of these rules, that is true if the rule under consideration is applicable on the active constraint of the given constraint sequence. The predicate takes the name of the rule, the input sequence $cs$, the output sequence $cs'$, input substitution $\sigma$, output substitution $\sigma'$, and the theory $\Th$ considered as arguments. For instance, we define $\msf{xor_r}$ as follows\footnote{$^{\frown}$ is the sequence concatenation operator.}:

\[
	\appl(\msf{xor_r},cs,cs',\sigma,\sigma',\Th)  
	\Lra
	(\ex m, T, t)
	\left(  
	\begin{array}{l}
	\act(m:T \cup t_1 \oplus \ldots \oplus t_n, cs) \wedge (\sigma' = \sigma) \wedge \\
	 (cs' = cs_< ^{\frown}[t_2 \oplus \ldots \oplus t_n:T, 
	 m:T \cup t_1]^{\frown} cs_>) 	 
	\end{array}
	 \right)
\]

We left out two important rules in the table, $\msf{un}$ and $\msf{ksub}$, that  change the attacker substitution through unification. We describe them next:

\[
	\appl(\msf{un},cs,cs',\sigma,\sigma',\Th)  
	\Lra
	(\ex m, T, t)
	\left(  
	\begin{array}{l}
	\act(m:T \cup t, cs) \wedge 
	 (cs' = cs_< \tau^{\frown}cs_>\tau) \wedge \\
	 (\sigma' = \sigma \cup \tau) \wedge (\tau \in U_E(\{ m \stackrel{?}{=}_E t  \}))
	\end{array}
	 \right)
\]

\[
	\appl(\msf{ksub},cs,cs',\sigma,\sigma',\Th)  
	\Lra
	(\ex m, T, t)
	\left(  
	\begin{array}{l}
	\act(m : T \cup [t]^{\to}_k, cs) \wedge \\
	 (cs' = cs_< \tau^{\frown}[m\tau : T\tau \cup [t]^{\to}_k\tau]^{\frown} 
	  cs_>\tau) \wedge \\
	 (\sigma' = \sigma \cup \tau) \wedge (\tau \in U_E(\{ k \stackrel{?}{=}_E \pk(\epsilon)  \}))
	\end{array}
	 \right)
\]

({\em Note}: $\epsilon$ is a constant of type $\Agent$ representing the name of the attacker).

We will say that a constraint sequence $cs'$ is a {\em child constraint sequence} of another sequence $cs$, if it can be obtained after applying some reduction rules on $cs$:  

\[
	\childseq(cs,cs',\Th) \Lra
	(\ex r_1, \ldots, r_n \in \Rules)
	\left(
		\begin{array}{l}
				\appl(r_1,cs,cs_1,\sigma,\sigma_1,\Th) \wedge \\
				\appl(r_2,cs_1,cs_2,\sigma_1,\sigma_2,\Th) \wedge \ldots \wedge \\
				\appl(r_n,cs_{n-1},cs',\sigma_{n-1},\sigma_n,\Th) 
		\end{array}								
	\right).
\]

We now define ``normal" constraint sequences, where the active constraint does not have sequences on the target or in the term set and has stand-alone variables in the term set (also recall that by definition, the target term of an active constraint is not a variable):

\[
	\normal(cs) \Lra
	\left(	
	 \begin{array}{c}
	 \act(m : T, cs)\wedge \\
	 (\nexists t_1,\ldots, t_n)( [t_1,\ldots,t_n] = m )\wedge  \\
	 ( (\fa t \in T)( (\nexists t_1,\ldots, t_n)( [t_1,\ldots,t_n] = t) ) \wedge \\
	 (\fa t \in T)(t \notin \mi{Vars}) )
	 \end{array}
	\right)
\]

Next, we will define a recursive function, $\normalize()$, that maps constraints to constraint sequences such that:

\begin{tabbing}
	$\normalize(m : T)$ \=  $=$ \= $[m : T]$,  if $\normal(m : T)$;\\
	\>  $=$	\>  $\normalize(t_1 : T)^{\frown}\ldots^{\frown}\normalize(t_n : T)$ if $m = [t_1,\ldots,t_n]$; \\
	\>  $=$ \>	$\normalize(m : T' \cup t_1 \cup \ldots \cup t_n)$ 
			if $T = T' \cup [t_1,\ldots,t_n]$.
\end{tabbing}

We will now overload this function to apply it on constraint sequences as well:

\begin{tabbing}
	$\normalize(cs)$ \= $=$  \=  $cs$, if $\normal(cs)$ \\
	\>  $=$ \> 	$cs_<^{\frown}\normalize(c)^{\frown}cs_>$, if $\act(c,cs)$.
\end{tabbing}

We define satisfiability of constraints as a predicate ``$\satisfiable$" which is true if there is a sequence of applicable rules which reduce a given normal constraint sequence $cs$ to a simple constraint sequence $cs_n$, in a theory $\Th$, resulting in a substitution $\sigma_n$: 

\begin{equation}\label{e.satisfiable}
\begin{array}{c}
\satisfiable(cs,\sigma_n,\Th) \Ra \\
	(\ex r_1,\ldots,r_n \in \Rules)
	\left(
	\begin{array}{l}
 		\appl(r_1,cs,cs_1,\{\}, \sigma_1, \Th) \wedge \\
  	\appl(r_2,cs'_1,cs_2,\sigma_1,\sigma_2, \Th) \wedge \ldots \wedge \\
    \appl(r_n,cs'_{n-1},cs_n,\sigma_{n-1},\sigma_n, \Th) \wedge \\
    \msf{simple}(cs_n) \wedge\\
   	(\fa i \in \{ 1,\ldots,n\})(cs'_i = \normalize(cs_i))
	\end{array}
\right).
\end{array}
\end{equation}

Notice the last clause which requires that every constraint sequence be normalized before any rule is applied, when checking for satisfiability.

This definition of satisfiability may seem unusual, especially for the puritans,  since satisfiability is usually defined using attacker capabilities as operators on sets of ground terms to generate each target on constraints.

However, it was proven in \cite{Chev04} that the decision procedure on which our definition is based, is sound and complete with respect to attacker capabilities on ground terms in the presence of the algebraic properties of \txor. Hence, we defined it directly in terms of the decision procedure, since that is what we will be using to prove our main theorem. We refer the interested reader to \cite{MS01} and \cite{Chev04} for more details on the underlying attacker operators, whose usage is equated to the decision procedure that we have used.

Note also that our definition only captures completeness of the decision procedure wrt satisfiability, not soundness, since that is the only aspect we need for our proofs in this paper.

\subsection{Security properties and attacks}\label{ss.security-prop}

Every security protocol is designed to achieve certain security goals such as  key establishment and authentication. Correspondingly, every execution of a protocol is expected to satisfy some related security properties. For instance, a key establishment protocol should not leak the key being established, which would be a violation of secrecy. It should also not lead an honest agent to exchange a key with an attacker, which would be a violation of both secrecy and authentication.

Our main result is general and is valid for any trace property such as secrecy,  that can be  tested by embedding the desired property into semi-bundles and then checking if constraint sequences from the semi-bundles are satisfiable:

\begin{definition}{{\em\bf [Secrecy]}}\label{d.secrecy}

A protocol is {\em secure for secrecy} in the theory $\Th$, if no constraint sequence from any semi-bundle of the protocol is satisfiable, after a strand with node that receives a secret constant is added to the semi-bundle. i.e., if $P$ is a protocol, then, 

\[
 (\nexists \mi{sec}, \msf{cs}, S)
	\left(
	\begin{array}{c}
	  \semibundle(S,P) \wedge \conseq(\msf{cs},S) \wedge \\
   (\msf{cs} = [\_ : \_, \ldots, \_ : T]) \wedge \\
			   (\mi{sec} \in \SecConstants(S)) \wedge \\
			   \satisfiable(\msf{cs}^{\frown}[\mi{sec} : T], \sigma, \Th) 
	\end{array}
	\right)  \Lra  \sfs(P,\Th).
\]
\end{definition}

\subsection{Main Requirement - $\munut$}\label{ss.MuNUT}

We now formulate our main requirement on protocol messages to prevent   multi-protocol attacks, namely $\munut$, in the $\SUA$ theory (an abbreviation for  $\STDUACUN$). The requirement is an extension of Guttman-Thayer's suggestion to make encrypted terms distinguishable across protocols, to include \txor{} as well.

We will first define a set $\XorTerms$ as:

\[
	\{ t \mid (\ex t_1,\ldots,t_n \in T(F,\Vars))(t_1 \oplus \ldots \oplus t_n = t)	\}.
\]

We will also define a function $\EnCp()$ that returns all the encrypted subterms of a set of terms. i.e., If $T$ is a set of terms, then, $\EnCp(T)$ is the set of all terms such that if $t$ belongs to the set, then $t$ must be a subterm of $T$ and is an encryption: 

\[	\EnCp(T) = \{ t \mid (\ex t', k')((t = [t']_{k'}) \wedge (t \in \SubTerms(T))) \}.	\]

Further, if $P$ is a protocol, then 

\[ \EnCp(P) = \{ t \mid t \in \EnCp(\SubTerms(P))	\}.  \]

We are now ready to state the main requirement formally:

\begin{definition}{{\em\bf [$\munut$]}}\label{d.munut}

	Two protocols $P_1$ and $P_2$ are $\munutsat$, i.e., $\munutsat(P_1,P_2)$ iff:

\begin{enumerate}

\item Encrypted subterms in both protocols are not $\std$-Unifiable after applying any substitutions to them: 

\[
		\begin{array}{c}
			(\fa t_1 \in \EnCp(P_1), t_2 \in \EnCp(P_2))((\nexists \sigma_1,\sigma_2)(t_1\sigma_1 =_{\std} t_2\sigma_2)).
		\end{array}
\]
	
\item Subterms of \txor-terms of one protocol (that are not \txor-terms themselves), are not $\std$-Unifiable with any subterms of \txor-terms of the other protocol (that are not \txor-terms as well):

\[
		\left(
				\begin{array}{l}
					\fa t_1 \oplus \ldots \oplus t_n \in \SubTerms(P_1), \\
					 t'_1 \oplus \ldots \oplus t'_n \in \SubTerms(P_2); t, t'
				\end{array}
		\right)
				\left(
				\begin{array}{c}
						(t \in \{ t_1, \ldots, t_n\}) \wedge 
						(t' \in \{t'_1, \ldots, t'_n\})  \\
						(t_1, \ldots, t_n, t'_1, \ldots, t'_n \notin \XorTerms) \wedge \\
						\Ra (\not\ex \sigma, \sigma')(t\sigma =_{\std} t'\sigma')
				\end{array}
				\right).
\]

\end{enumerate}

\end{definition}

The first requirement is the same as Guttman-Thayer suggestion. The second requirement extends it to the case of \txor-terms, which is our stated extension in this paper.

The $\msf{NSL}_{\oplus}$ protocol can be transformed to suit this requirement by tagging its encrypted messages as follows:

\begin{center}
\begin{tabular}{ll}

{\bf Msg 1.} $A \to B : [\nslxor, N_A, A]_{\mathit{pk}(B)}$ \\

{\bf Msg 2.} $B \to A : [\nslxor, [\nslxor,N_A] \oplus [\nslxor,B], N_B ]_{\mathit{pk}(A)}$ \\

{\bf Msg 3.} $A \to B : [\nslxor, N_B]_{pk(B)}$\\

\end{tabular}
\end{center}

The constant ``$\nslxor$" inside the encryptions can be encoded using some suitable bit-encoding when the protocol is implemented. Obviously, other protocols must have their encrypted subterms start with the names of those protocols.

\section{A Lynchpin Lemma}\label{s.a-lemma}

In this section, we provide a useful lemma that is the lynchpin in achieving our main result. We prove in the lemma that, if we follow BSCA for $(\SUA)$-UPs that do not have \txor{} terms with variables, their \tacun{} subproblems will have only constants as subterms. 

%Consequently, we will end up in an empty set of substitutions returned by the {\sf ACUN} unification algorithm for the {\sf ACUN} unification problems, even when the \txor{}  terms are equal in the {\sf ACUN} theory.

\begin{lemma}{  {\em\bf [$\acun$ UPs have only constants]}}\label{l.acun-nosub}

Let $\Gamma = \{m \stackrel{?}{=}_{\SUA} t\}$ be a $(\SUA)$-UP that is $(\SUA)$-Unifiable, and where no subterm of $m$ or $t$ is an \txor{} term with free variables\footnote{$\mathbb{N}$ is the set of natural numbers.}:

\[
		(\fa x)
		\left(
			\begin{array}{c}
				( (x \sqss m) \vee (x \sqss t) )  \wedge
				(n \in \mathbb{N}) \wedge \\
				(x = x_1 \oplus \ldots \oplus x_n)
			\end{array}
			\Ra
			(\fa i \in \{1,\ldots,n\})(x_i \notin \Vars)
		\right).
\]

Then,

\[
(\fa m' \stackrel{?}{=}_{\acun} t' \in \Gamma_{5.2};y)
	\left(
	\begin{array}{c}
		\left(
		\begin{array}{c}
			((y \sqsubset m') \vee (y \sqsubset t')) \wedge \\
			(m' =_{\acun} t')
		\end{array}
		\right)
	\Ra
		(y \in \Constants)
	\end{array}
	\right).
\]

\end{lemma}

\begin{proof}

Please see Appendix~\ref{s.proofs}, Lemma~\ref{la.acun-nosub}.

\end{proof}

\section{Main result - $\mu$-{\sf NUT} prevents multi-protocol attacks}\label{s.main-result}

We will now prove that $\munutsat$ protocols are not susceptible to multi-protocol attacks. 

The idea is to show that if a protocol is secure in isolation, then it is in combination with other protocols with whom it is $\munutsat$.

To show this, we will achieve a contradiction by attempting to prove the contrapositive. i.e., if there is a breach of secrecy for a protocol in combination with another protocol with which it is $\munutsat$, then it must also have a breach of secrecy in isolation.

We assume that the reader is familiar with BSCA (detailed description in Appendix~\ref{s.BSCA}).

\begin{theorem}\label{t.multi-prot}

	If a protocol is secure for secrecy, then it remains so in combination with any other protocol with which it is  $\munutsat$.

\end{theorem}

\begin{proof}

Suppose  $P_1$  is a protocol that is secure for secrecy in isolation in the $\SUA$ theory. i.e., $\sfs(P_1, \SUA)$. Consider another protocol  $P_2$  such that,  $\munutsat(P_1,P_2)$.  Let,  $S_1$  and  $S_2$  be two semi-bundles from  $P_1$  and  $P_2$  respectively: 

\[ \semibundle(S_1,P_1)	\wedge  \semibundle(S_2,P_2).  \] 

Consider a constraint sequence  $\mi{combcs}$  from  $S_{\mi{comb}} = S_1 \cup S_2$.  i.e., 

\[ \conseq(\combcs, S_{\mi{comb}}).  \] 

Consider another constraint sequence  $\isocs$, where, 

{\bf (a)} Targets in  $\combcs$  are targets in  $\isocs$  if the targets belong to  $S_1$:

\begin{equation}\label{e.same-targets}
		(\fa m : \_~\msf{in}~\combcs)( (m \in \Terms(S_1)) \Ra (m : \_~\msf{in}~ \isocs) ).
\end{equation}
 
{\bf (b)} Term sets in  $\combcs$  are term sets in  $\isocs$  but without terms from  $S_2$:
	
\begin{equation}\label{e.same-termsets}
		\left(
			\begin{array}{l}
				\fa m_1 : T_1, \\
				    m_2 : T_2 ~\msf{in} ~\combcs
			\end{array}
		\right)
		\left(
			\begin{array}{c}
				m_1 : T_1 \prec_{\combcs} m_2 : T_2 \\
				\Ra \\
				(\ex T'_1, T'_2)
				\left(
					\begin{array}{c}
							(m_1 : T'_1 \prec_{\isocs} m_1 : T'_2) \wedge \\
							(T'_1 = T_1 \setminus T''_1) \wedge (T'_2 = T_2 \setminus 																														T''_2) \\
							(\fa t \in T''_1 \cup T''_2)(t \in \SubTerms(S_2))
					\end{array}
				\right)	
			\end{array}
		\right).
\end{equation}

Then, from Def.~\ref{d.constraints} ({\bf Constraints}) we have: $\conseq(\isocs,S_1)$.

Suppose  $\combcs$  and  $\isocs$  are normalized. To achieve a contradiction, let there be a violation of secrecy in  $S_{\mi{comb}}$  s.t.  $\combcs$  is satisfiable after an artificial constraint with a secret constant of  $S_1$, say $\mi{sec}$,  is added to it:

\begin{equation}
		(\combcs = [ \_  :  \_, \ldots, \_  :  T ]) \wedge 								 			\satisfiable(\combcs^{\frown}[\mi{sec}  :  T], \_, \SUA). 
\end{equation}

Suppose $[r_1, \ldots, r_n] = R$, such that $r_1,\ldots,r_n \in \Rules$. Then, from the definition of satisfiability (\ref{e.satisfiable}), using  $R$,   say we have:

\begin{equation}\label{e.comb-satisfiable}
\left(
	\begin{array}{l}
	 (\combcs = [\_ : \_, \ldots, \_ : T]) \wedge \\
	 \appl(r_1,\combcs^{\frown}[\mi{sec} : T],\combcs_1,\{\}, \sigma_1, \SUA)  \wedge \\
  	\appl(r_2,\combcs'_1,\combcs_2,\sigma_1,\sigma_2, \SUA)  \wedge  \ldots \wedge \\
    \appl(r_n,\combcs'_{n-1},\combcs_n,\sigma_{n-1},\sigma_n, \SUA)  \wedge \\
    \simple(\combcs_n)  \wedge 
   (\fa i \in \{  1,\ldots,n \})(\combcs'_i = \normalize(\combcs_i))  
\end{array}
\right).
\end{equation}

From their descriptions, every rule in $\Rules$ adds subterms of existing terms (if any) in the target or term set of the active constraint:

\begin{equation}\label{e.rules-add-subterms}
\left(
	\begin{array}{c}
		 \appl(\_,cs,cs',\_,\_,\_) \wedge	\act(m : T, cs) \wedge \\
	   \act(m' : T', cs')  \wedge (x \in T' \cup \{m'\}) 
\end{array}
\right)
  	    \Ra 
  	  (x \in \SubTerms(T \cup \{m\})).
\end{equation}

Since every $\combcs'_i$ ($i = 1$ to $n$) in (\ref{e.comb-satisfiable}) is normalized, and since $P_1$ and $P_2$ are $\munutsat$, we have:

\begin{equation}\label{e.xor-terms-no-vars}
	(\fa i \in \{1,\ldots,n\}; \ex p \in \mathbb{N}; t_1, \ldots, t_p)
		\left(
			\begin{array}{c}
				\act(m : T, \combcs'_i) \wedge \\
				(t_1 \oplus \ldots \oplus t_p \in T \cup \{m\}) \Ra \\
				(\fa j \in \{1,\ldots,p\})(t_j \notin \Vars)
			\end{array}
		\right).
\end{equation}

Suppose  $\chcombcs$  is a normal, child constraint sequence of  $\combcs$  and  $\chisocs$  is a normal, child constraint sequence of  $\isocs$. 

Now all the rules in  $\Rules$  are applicable on the target of the active constraint of  $\chisocs$,  if they were on  $\chcombcs$,  provided they are applied on a term of  $S_1$:  

\begin{equation}\label{e.target-rules-apply-equally}
\begin{array}{l}
(\fa r \in \Rules) 
	\left(
		\begin{array}{c}
			\appl(r,\chcombcs,\chcombcs',\_,\_, \SUA) \wedge \\
			\act(m : \_,\chcombcs) \wedge  
			\act(m' : \_,\chcombcs') 	\wedge \\
			\act(m : \_,\chisocs)
		\end{array}
	\right)
	\Ra  \\
	\left(
		\begin{array}{c}
			\appl(r,\chisocs,\chisocs',\_,\_, \SUA) \wedge 
			\act(m' : \_,\chisocs')
		\end{array}
	\right).
\end{array}
\end{equation}

Similarly, all rules that are applicable on a term in the term set of the active constraint in  $\chcombcs$, say $c$,  are also applicable on the same term of the active constraint in  $\chisocs$, say $c'$ (provided the term exists in the term set of $c'$, which it does from (\ref{e.same-termsets}) and (\ref{e.rules-add-subterms})): \\

\begin{equation}\label{e.termset-rules-apply-equally}
\begin{array}{l}
(\fa r \in \Rules) 
	\left(
		\begin{array}{c}
			\appl(r,\chcombcs,\chcombcs',\_,\_, \SUA) \wedge \\
			\act(\_ : \_ \cup t,\chcombcs) \wedge 
			\act(\_ : \_ \cup T' ,\chcombcs') \wedge \\
			\act(\_ : \_ \cup t,\chisocs)
		\end{array}
	\right)
	\Ra \\
	\left(
		\begin{array}{c}
			\appl(r,\chisocs,\chisocs',\_,\_, \SUA) \wedge 
			\act(\_ : \_ \cup T',\chisocs')
		\end{array}
	\right).
\end{array}
\end{equation}

$\msf{un}$  and  $\msf{ksub}$ are the only rules that affect the attacker substitution.  We will show that these are equally applicable on $\chcombcs$ and $\chisocs$ as well. Suppose:

\begin{itemize}

	\item $\Gamma = \{ m \stackrel{?}{=}_{\SUA} t \}$, is a $(\SUA)$-{\sf UP} 					and suppose  $m = m' \subcomb$, $t = t'\subcomb$, where  $m' \in \SubTerms(S_1)$;

	\item Variables in $\subcomb$ are substituted with terms from the same 				semi-bundle: 
	
				\begin{equation}\label{e.disj-subs}
					(\fa x/X \in \subcomb)((\ex i \in \{1,2\})(x,X \in \SubTerms(S_i))).
				\end{equation}

	\item $\Gamma$  is $(\SUA)$-Unifiable. 

\end{itemize}

Let $\tau \in U_{\SUA}(\Gamma)$ and let $A_{\Th}$ denote a $\Th$-UA. Using Def.~\ref{d.Combined-Unifier} ({\bf Combined Unifier}),  say we have that $\tau \in \tau_{\std} \odot \tau_{\acun}$  where  $\tau_{\std} \in A_{\std}(\Gamma_{5.1})$  and  $\tau_{\acun} \in A_{\acun}(\Gamma_{5.2})$.

Now from BSCA, if  $m_1 \stackrel{?}{=}_{\std} t_1 \in \Gamma_{5.1}$, and $\rho \in U_{\std}(m_1 \stackrel{?}{=}_{\std} t_1)$, then we have the following cases:

\paragraph*{Variables.} If  $m_1$, and/or  $t_1$  are variables, from (\ref{e.xor-terms-no-vars}) and BSCA, they are necessarily new i.e., $m_1, t_1 \in \Vars \setminus \Vars(\Gamma)$ (unless  $m$  and  $t$  are variables, which they are not, since  $\chcombcs$  is normal). Hence, there are no new substitutions in $\rho$ to $\Vars(\Gamma)$ in this case.

\paragraph*{Constants.} If $m_1 \in \Constants(S_1)$, again from BSCA, $t_1$ cannot belong to $\Vars$, and it must be a constant. If $m_1$ is a fresh constant of $S_1$, then $t_1$ must also belong to $S_1$ from the freshness assumption (\ref{a.freshness}) and (\ref{e.disj-subs}), and if $m_1$ is not fresh, $t_1$ could belong to either $\SubTerms(S_1)$ or $\iik$ from Assumption~\ref{a.IIK2}. Further, $\rho = \{ \}$.
	
\paragraph*{Public Keys.} If $m_1 = \pk(\_)$, then $t_1$ must be some $\pk(\_)$  as well. From BSCA, $m_1$ cannot be such that  $\penc{\_}{m_1} \sqsubset m$. Further, there cannot be an \txor{} term, say $\ldots \oplus m_1 \oplus \ldots$ that is a subterm of $m$, from $\munut{}$ Condition~2. The only other possibility is that $m = m_1$. In that case, $t$ must also equal $t_1$, whence, $t$ can belong to $\iik$ from assumption~\ref{a.IIK1} (Intruder possesses all public-keys). Hence, $(\fa x/X \in \rho)((\ex i \in \{1,2\})(x, X \in \SubTerms(S_i)))$. 

%\paragraph*{Shared keys.} $m_1$ cannot be a long-term shared-key; i.e., $m_1 \neq \sh(\_,\_)$, since from Assumption~\ref{a.LTKeys}, they do not appear as interms and from the definition of $\Gamma_{5.1}$, $m_1$ is necessarily an interm.
	
\paragraph*{Encrypted Subterms.} 	Suppose $m_1 = m_{11}\subcomb$, $t_1 = t_{11}\subcomb$, $m_{11}, t_{11} \in \EnCp(S_1 \cup S_2)$.
Then, from $\mu$-{\sf NUT} Condition~1 and (\ref{e.rules-add-subterms}), we have, $m_{11}, t_{11} \in 	\EnCp(S_i)$, where $i \in \{1,2\}$. Hence, $(\fa x/X \in \rho)((\ex i \in \{1,2\})(x,X \in \SubTerms(S_i)))$. 			

\paragraph*{Sequences.} If $m_1$ is a sequence, either $m$ must be a sequence, or there must be some $\ldots \oplus m_1 \oplus \ldots$ belonging to $\SubTerms(\{m,t\})$, from BSCA. But $m$ and $t$ cannot be sequences, since $\chcombcs$ is normal. Hence, by $\munut{}$ Condition~2 and (\ref{e.rules-add-subterms}), $m_1, t_1 \in \SubTerms(S_i)\subcomb$, $i \in \{1,2\}$ and $(\fa x/X \in \rho)((\ex i \in \{1,2\})(x,X \in \SubTerms(S_i)))$. 			

In summary, we make the following observations about problems in $\Gamma_{5.1}$.

If $m_1$ is an instantiation of a subterm in $S_1$, then so is $t_1$, or $t_1$ belongs to $\iik$:

\begin{equation}\label{e.t1-belongs-to-S1}
		(\fa m_1 \stackrel{?}{=}_{\std} t_1 \in \Gamma_{5.1})(m_1 \in \SubTerms(S_1)\subcomb \Ra t_1 \in \SubTerms(S_1)\subcomb \cup \iik).
\end{equation}

Every substitution in $\tau_{\std}$ has both its term and variable from the same semi-bundle:

\begin{equation}\label{e.tau-std-subs}
	(\fa x/X \in \tau_{\std})((\ex i \in \{1,2\})(x,X \in \SubTerms(S_i))).	
\end{equation}

Now consider the UPs in $\Gamma_{5.2}$. Applying (\ref{e.xor-terms-no-vars}) into  Lemma~\ref{l.acun-nosub}, we have that $\tau_{\acun} = \{ \}$. Combining this with (\ref{e.tau-std-subs}), we have: 

\begin{equation}\label{e.tau-pure} (\fa x/X \in \tau)((\ex i \in \{1,2\})(x,X \in \SubTerms(S_i)\subcomb)). 
\end{equation}

Suppose $m = m_1 \oplus \ldots \oplus m_p$ and $t = t_1 \oplus \ldots \oplus t_q$; $p, q \ge 1$, $x = m\tau$, $y = t\tau$ and $m'' =_{\SUA} x$ where $m'' = m'_1 \oplus \ldots \oplus m'_{p'}$, s.t. $(\fa i, j \in \{1,\ldots,p'\})(i \neq j \Ra m'_i\tau \neq_{\SUA} m'_j \tau)$ and $t'' =_{\SUA} y$, where $t'' = t'_1 \oplus \ldots \oplus t'_{q'}$, s.t. $(\fa i,j \in \{1,\ldots,q'\})(i \neq j \Ra t'_i\tau \neq_{\SUA} t'_j\tau)$. Informally, this means that, no two terms in $\{m'_1,\ldots,m'_{p'}\}$ or $\{t'_1,\ldots,t'_{q'}\}$ can be cancelled.

Suppose $\Gamma\psi = \Gamma_{5.1}$, where $\psi$ is a set of substitutions. Then, $m\tau =_{\SUA} t\tau$ implies, $(\fa i \in \{1,\ldots,p'\})((\ex j \in \{1,\ldots,q'\})(m'_i\tau\psi =_{\std} t'_j\tau\psi))$ with $p' = q'$. 
From $(\ref{e.t1-belongs-to-S1})$, this means that $m \in \SubTerms(S_1)\subcomb$ implies, $t$ also belongs to $\SubTerms(S_1)\subcomb$ or $\iik$.

Now since $\Vars(m') \cup \Vars(t') \subset \Vars(S_1)$, we have, $m'\subcomb = m'\subiso$, and $t'\subcomb = t'\subiso$, where $\subcomb = \subiso \cup \{x/X \mid x, X \in \SubTerms(S_2) \}$. Combining this with (\ref{e.tau-pure}), we have that, $m'\subcomb\tau =_{\SUA} t'\subcomb\tau \Ra m'\subiso\tau =_{\SUA} t'\subiso\tau$.

Combining these with (\ref{e.same-targets}) and (\ref{e.same-termsets}), we can now write: 
		\begin{equation}\label{e.un-applies-equally}
			(\fa \chcombcs, \chisocs)
			\left(
			\begin{array}{l}
				\childseq(\chcombcs,\combcs,\SUA) \wedge \\
				\childseq(\chisocs,\isocs,\SUA) \wedge \\
				\appl(\msf{un},\chcombcs,\chcombcs',\subcomb,\subcomb',\SUA) 	\Ra \\
				\appl(\msf{un},\chisocs,\chisocs',\subiso,\subiso',\SUA) 
			\end{array}
			\right).
		\end{equation}

where, the active constraint in $\chcombcs$ and $\chisocs$ only differ in the   term sets:

\[
		\left(  
	\begin{array}{l}
		\act(m : \_ \cup t, \combcs) \wedge 	\act(m : \_ \cup t, \isocs) \wedge \\
		 (\combcs' = \combcs_<\tau^{\frown}
	 	\combcs_>\tau)  \wedge  	 (\isocs' = \isocs_<\tau^{\frown} 
	 	\isocs_>\tau) \wedge \\
	 	(\subcomb' = \subcomb \cup \tau) \wedge 	 (\subiso' = \subiso \cup \tau) 			\wedge 	 	(\tau \in \tmgu{(\SUA)}(\tuple{m}{t}))
	\end{array}
	 \right)
\]

Finally, we can combine, (\ref{e.comb-satisfiable}),  (\ref{e.target-rules-apply-equally}), (\ref{e.termset-rules-apply-equally}), and (\ref{e.un-applies-equally}) to infer:

\begin{equation}\label{e.satisfiability-achieved-for-cs1}
	\left(
 \begin{array}{l}
	(\isocs = \lan \_ : \_, \ldots, \_ : T \ran) \wedge
  \appl(r_1,\isocs^{\frown}[\mi{sec} : T],\isocs_1,\{\}, \sigma_1, \SUA) \wedge \\
  \appl(r_2,\isocs'_1,\isocs_2,\sigma_1,\sigma_2, \SUA) \wedge \ldots \wedge \\
  \appl(r_p,\isocs'_{p-1},\isocs_p,\sigma_{p-1},\sigma_p,\SUA) \wedge \\
  \msf{simple}(\isocs_p) \wedge 
  (\fa i \in \{ 1,\ldots,p\})(\isocs'_i = \normalize(\isocs_i))
\end{array}
\right).
\end{equation}

\noindent
where $[r_1,\ldots,r_p]$ is a subsequence\footnote{$s'$ is a {\em subsequence} of a sequence $s$, if $s = \_^{\frown}s'^{\frown}\_$.} of $R$ (defined in \ref{e.comb-satisfiable}).

This in turn implies $\satisfiable(\isocs^{\frown}\mi{sec} : T,\sigma_p,\SUA)$ from the definition of satisfiability.

We can then combine this with the fact that $S_1$ is a semi-bundle of $P_1$, and $\isocs$ is a constraint sequence of $S_1$ and conclude:

\[
	\begin{array}{c}
	  \semibundle(S_1,P_1) \wedge \conseq(\isocs,S_1) \wedge 
	  (\isocs = [\_ : \_, \ldots, \_ : T]) \wedge \\
	  \satisfiable(\isocs^{\frown}[\mi{sec} : T], \sigma_p,\SUA). 
	\end{array}
\]

But from Definition \ref{d.secrecy} ({\bf Secrecy}), this implies, $\neg \sfs(P_1,\SUA)$, a contradiction to the hypothesis. Hence, $P_1$ is always secure for secrecy in the $(\SUA)$ theory, in combination with $P_2$ with which it is $\munutsat$.

\end{proof}

\section{Conclusion}\label{s.conclusion}

In this paper, we provided a formal proof that tagging to ensure non-unifiability of distinct encryptions prevents multi-protocol attacks under the {\sf ACUN} properties induced by the Exclusive-OR operator. We will now discuss some prospects for future work and related work.

\subsection{Future work}\label{ss.future-work}

Other equational theories can be handled in the same way as the \tacun{} theory: When we use BSCA, the UPs for them ($\Gamma_{5.2}$) will only have constants as subterms. Hence, unifiers only from the algorithms for standard theory problems need to be considered for $\munutsat{}$ protocols. Of course, this reasoning has to be given within a symbolic constraint solving model that takes the additional equational theories into account (the model we used, adapted from~\cite{Chev04}, was tailored to accommodate only \tacun).

We achieved our main result specifically for secrecy. The reason for this was that, in order to prove that attacks exist in isolation if there did in combination, we had to have a precise definition as to what an ``attack" was to begin with. However, other properties such as authentication and observational equivalence can be considered on a case-by-case basis, with a similar proof pattern. 

At the core of our proofs is the use of BSCA for combined theory unification. However, BSCA is applicable only for disjoint theories that do not share any operators. For instance, the algorithm cannot consider equations of the form, $[ a, b ]  \oplus  [ c, d ]   =   [ a \oplus c, b \oplus d ]$. 

We plan to expand our proofs to include such equations in future, possibly with the help of new unification algorithms~\cite{ALLNR09}.

%\paragraph*{Funding.} This work funded in part by a doctoral SEED grant by the Graduate School at Dakota State University. I am particularly grateful to Dean~Tom~Halverson (college of BIS) and Dean~Omar~El-Gayar (college of graduate studies and research) for their continued support for my research.

\subsection{Related work}\label{ss.related-work}

To the best of our knowledge, the consideration of algebraic
properties and/or equational theories for protocol independence is unchartered waters.

A study of multi-protocol attacks with the perfect encryption 
assumption relaxed was first reported by Malladi et al. in~\cite{MAM02} 
through ``multi-protocol guessing attacks" on password protocols. 
Delaune et al. proved that these can be prevented by tagging 
in~\cite{DKR-csf08}.

The original work of Guttman et al. in
~\cite{TG00} assumed that protocols have no type-flaw attacks when
they proved that tagging to ensure disjoint encryption prevents multi-protocol
attacks. But a recent work by Guttman seems to relax that assumption
~\cite{Guttman09}. Both~\cite{TG00} and~\cite{Guttman09} use the strand space model~\cite{THG98}. Our protocol model in this paper is also based on strand spaces, but the penetrator actions are modeled as symbolic reduction rules in the constraint solving algorithm of~\cite{Chev04,MS01}, as opposed to penetrator strands in~\cite{THG98}. 
Cortier-Delaune also prove that multi-protocol attacks can be prevented with tagging, which is
slightly different from~\cite{TG00} and considers composed/non-atomic keys~\cite{CD09}. They too seem to use the constraints model as their protocol framework.

In~\cite{CM07}, we prove the decidability of tagged protocols that use \txor{} with the underlying framework of~\cite{Chev04} which extends~\cite{MS01} with \txor. That work is similar to our proofs, since we too used the same framework (\cite{Chev04}). Further, we use BSCA~\cite{BS96} as a core aspect of this paper along the lines of~\cite{CM07}. Recently, we used a similar proof pattern to prove that tagging prevents type-flaw attacks under \txor{} and most likely under other equational theories in~\cite{ML09}. Lemma~\ref{l.acun-nosub} in the current paper was also the lynchpin in~\cite{ML09}.

In~\cite{KustersT08}, Kuesters and Truderung showed that the verification of protocols that use the \txor{} operator can be reduced to verification in a free term algebra, for a special class of protocols called $\oplus$-linear protocols\footnote{Kuesters-Truderung define a term to be $\oplus$-linear if for each of its subterms of the form $s \oplus t$, either $t$ or $s$ is ground.}, so that {\sf ProVerif} can be used for verification. Chen et al. recently report some extensions to Kuesters-Truderung's work~\cite{CDP09}.

These results have a similarity with ours, in the sense that we too show that the algebraic properties of \txor{} have no effect when some of the messages are modified. However, we believe that our result is more general than these, since any protocol can be tagged to satisfy our requirements, but not necessarily $\oplus$-linearity.

\paragraph*{Acknowledgments.} I am thankful to Yannick Chevalier for explaining his protocol model in~\cite{Chev04}, Pascal Lafourcade for many useful remarks and the anonymous reviewers for their helpful suggestions.

%\renewcommand{\baselinestretch}{0.98}
%\bibliographystyle{apalike}
%{\small
%\bibliography{SECRYPT09bib}}
%\renewcommand{\baselinestretch}{1}

{\small 
	\bibliographystyle{splncs}
	\bibliography{FCS10}
}

%\tableofcontents

\newpage
\appendix
\section{Bader \& Schulz Combined Theory Unification Algorithm (BSCA)}\label{s.BSCA}

We will now consider how two UAs for two disjoint theories $\Th_1$ and $\Th_2$ respectively, may be combined to output the unifiers for UPs made using operators from $\Th_1 \cup \Th_2$ using Baader \& Schulz Combination Algorithm (BSCA)~\cite{BS96}.

We first need some definitions. Suppose $F$ is a signature for a set of identities $E$ and let $\Th$ denote the theory $=_E$. Then, a term is pure wrt $\Th$ iff every subterm of it is an $F$-term. i.e.,

\[	\pure(t,\Th) \Lra (\fa t' \sqsubset t)((\ex f \in F)(t' = f(\_,\ldots,\_))).	\]

We define a predicate $\msf{ast}$ (alien subterm) on terms such that, a term $t'$ is an alien subterm of another term $t$ wrt the theory $\Th$, if it is a subterm of $t$, but is not pure wrt $\Th$:

\[   (\fa t, t', \Th)( \msf{ast}(t',t,\Th) \Lra (t' \sqsubset t) \wedge \neg\pure(t',\Th) ).   \]

For instance, $[1, n_a \oplus B, A]^{\to}_{\mi{pk}(B)}$ has $n_a \oplus B$ as an alien subterm with respect to the theory $\std$.

We will use the following $(\std \cup \acun)$-UP as our running example\footnote{We omit the superscript $\to$ on encrypted terms in this problem, since they obviously use only asymmetric encryption.}:

\[ \left\{ [1,n_a]_{\mi{pk}(B)} \stackrel{?}{=}_{\sua} [1,N_B]_{\mi{pk}(a)}   \oplus   [2,A] \oplus  [2,b] \right\}. \]

BSCA takes as input a $(\Th_1 \cup \Th_2)$-UP, say $\Gamma$, and applies some transformations on them to derive $\Gamma_{5.1}$ and $\Gamma_{5.2}$ that are $\Th_1$-UP and $\Th_2$-UP respectively. 

\subsubsection*{Step 1 (Purify terms)} BSCA first ``purifies" the given set of $(\Th = \Th_1 \cup \Th_2)$-UP, $\Gamma$, into a new set of problems $\Gamma_1$, such that, all the terms are pure wrt $\Th_1$ or $\Th_2$.

If our running example was $\Gamma$, then, the set of problems in $\Gamma_1$ are $W \stackrel{?}{=}_{\std} [1,n_a]_{\mi{pk}(B)}$, $X  \stackrel{?}{=}_{\std}  [1,N_B]_{\mi{pk}(a)}, Y  \stackrel{?}{=}_{\std} [2,A]$, $Z  \stackrel{?}{=}_{\std}  [2,b]$, and $W \stackrel{?}{=}_{\acun} X \oplus Y \oplus Z$, where $W, X, Y, Z$ are obviously new variables that did not exist in $\Gamma$.

\subsubsection*{Step 2. (Purify problems)} Next, BSCA purifies $\Gamma_1$ into $\Gamma_2$ such that, every problem in $\Gamma_2$ has both terms pure wrt the same theory.

For our example problem, this step can be skipped since all the problems in $\Gamma_1$ already have both their terms purely from the same theory ($\std$ or $\acun$)). 

\subsubsection*{Step 3. (Variable identification)} Next, BSCA partitions $\Vars(\Gamma_2)$ into a partition $\VarIdP$ such that, each variable in $\Gamma_2$ is replaced with a representative from the same equivalence class in $\VarIdP$. The result is $\Gamma_3$.

In our example problem, one set of values for $\VarIdP$ can be 
\[ \left\{ \{A\},\{B\},\{N_B\}, \{W\},\{X\},\{Y,Z\} \right\}. \]

\subsubsection*{Step 4. (Split the problem)} The next step of BSCA is to split $\Gamma_3$ into two UPs $\Gamma_{4.1}$ and $\Gamma_{4.2}$ such that, each of them has every problem with terms from the same theory, $\Th_1$ or $\Th_2$.
 
Following this in our example,

\[  \Gamma_{4.1} = \left\{ W \stackrel{?}{=}_{\std} [1,n_a]_{\mi{pk}(B)}, X  \stackrel{?}{=}_{\std} [1,N_B]_{\mi{pk}(a)}, Y \stackrel{?}{=}_{\std} [2,A], Z \stackrel{?}{=}_{\std} [2,b]  \right\}, \] and 

\[ \Gamma_{4.2} = \left\{ W \stackrel{?}{=}_{\acun} \xorseq{X}{Y}{Y} \right\}. \]

\subsubsection*{Step 5. (Solve systems)} The penultimate step of BSCA is to partition all the variables in $\Gamma_3$ into a size of two: Let $p = \{ V_1, V_2 \}$ is a partition of $\Vars(\Gamma_3)$. Then, the earlier problems ($\Gamma_{4.1}$, $\Gamma_{4.2}$) are further split such that, all the variables in one set of the partition are replaced with new constants in the other set and  vice-versa. The resulting sets are $\Gamma_{5.1}$ and $\Gamma_{5.2}$.

In our sample problem, we can form $\{ V_1, V_2 \}$ as $\{ \Vars(\Gamma_3), \{\} \}$. i.e., we choose that all the variables in problems of $\Gamma_{5.2}$ be replaced with new constants. This is required to find the unifier for the problem (this is the partition that will successfully find a unifier).

So $\Gamma_{5.1}$ stays the same as $\Gamma_{4.1}$, but $\Gamma_{5.2}$ is changed to 

\[	\Gamma_{5.2} = \Gamma_{4.2} \beta  
=	 \left\{ W \stackrel{?}{=}_{\acun} \xorseq{X}{Y}{Y} \right\} \beta  =  \left\{ w \stackrel{?}{=}_{\acun} \xorseq{x}{y}{y} \right\}.
\]

 i.e., $\beta = \left\{ w/W, x/X, y/Y \right\}$, where, $w, x, y$ are constants, which obviously did not appear in $\Gamma_{5.1}$. 

\subsubsection*{Step 6. (Combine unifiers)} The final step of BSCA is to combine the unifiers for $\Gamma_{5.1}$ and $\Gamma_{5.2}$, obtained using $A_{\Th_1}$ and $A_{\Th_2}$:

\begin{definition}{\em\bf [Combined Unifier]}\label{d.Combined-Unifier}

Let $\Gamma$ be a $\Th$-UP where $(\Th_1 \cup \Th_2) = \Th$. Let $\sigma_i \in A_{\Th_i}(\Gamma_{5.i})$, $i \in \{1,2\}$ and let $V_i = \Vars(\Gamma_{5.i})$, $i \in \{1, 2 \}$.

Suppose `$<$' is a linear order on $\Vars(\Gamma)$ such that $Y < X$ if $X$ is not a subterm of an instantiation of $Y$:

	\[	(\fa X, Y \in \Vars(\Gamma))((Y < X) \Ra (\not\ex \sigma)(X \sqsubset Y\sigma)).  \]

Let $\msf{least}(X,T,<)$ be defined as the minimal element of set $T$, when ordered linearly by the relation `$<$'. i.e., 

\[	\msf{least}(X,T,<)	\Lra	(\fa Y \in T)((Y \neq X) \Ra (X < Y)).	\]

Then, the combined UA for $\Gamma$, namely $A_{\Th_1 \cup \Th_2}$, is defined such that,

\[	A_{\Th_1\cup \Th_2}(\Gamma) = \{ \sigma \mid (\ex \sigma_1,\sigma_2)((\sigma = \sigma_1 																					\odot \sigma_2) \wedge (\sigma_1 \in A_{\Th_1}(\Gamma_{5.1})) \wedge (\sigma_2 \in A_{\Th_2}(\Gamma_{5.2}))) \}.
\] 

\noindent
where, if $\sigma = \sigma_1 \odot \sigma_2$, then,

\begin{itemize}

\item The substitution in $\sigma$ for the least variable in $V_1$ and $V_2$ is  from $\sigma_1$ and $\sigma_2$ respectively: \\

$(\fa i \in \{ 1, 2 \})( (X \in V_i) \wedge \msf{least}(X, \Vars(\Gamma), <)  \Rightarrow (X\sigma = X\sigma_i))$; and \\

\item For all other variables $X$, where each $Y$ with $Y < X$ has a substitution already defined, define
$X\sigma = X\sigma_i\sigma$ $(i \in \{1,2\})$: \\

$(\fa i \in \{ 1, 2 \})( (\fa X \in V_i)( (\fa Y)( (Y < X) \wedge (\ex Z)(Z/Y \in \sigma) )) \Rightarrow (X\sigma = X \sigma_i \sigma))$.

\end{itemize}

\end{definition}
\section{Proofs}\label{s.proofs}

The following lemma concerns combined unification problems involving \tstd{} and \tacun{} theories. We prove that, if we follow Bader \& Schulz approach for finding unifiers for these problems, \tacun{} subproblems will have only constants as subterms. Consequently, we will end up in an empty set of substitutions returned by the \tacun{} UA for the \tacun{} UPs, even when the \txor{}  terms are equal in the \tacun{} theory.

\begin{lemma}{  {\em\bf [$\acun$ UPs have only constants]}}\label{la.acun-nosub}

Let $\Gamma = \{m \stackrel{?}{=}_{\SUA} t\}$ be a $(\SUA)$-UP that is $(\SUA)$-Unifiable, and where no subterm of $m$ or $t$ is an \txor{} term with free variables\footnote{$\mathbb{N}$ is the set of natural numbers.}:

\[
		(\fa x)
		\left(
			\begin{array}{c}
				( (x \sqss m) \vee (x \sqss t) )  \wedge
				(n \in \mathbb{N}) \wedge \\
				(x = x_1 \oplus \ldots \oplus x_n)
			\end{array}
			\Ra
			(\fa i \in \{1,\ldots,n\})(x_i \notin \Vars)
		\right).
\]

Then,

\[
(\fa m' \stackrel{?}{=}_{\acun} t' \in \Gamma_{5.2};y)
	\left(
	\begin{array}{c}
		\left(
		\begin{array}{c}
			((y \sqsubset m') \vee (y \sqsubset t')) \wedge \\
			(m' =_{\acun} t')
		\end{array}
		\right)
	\Ra
		(y \in \Constants)
	\end{array}
	\right).
\]

\end{lemma}

\begin{proof}

	Let   $\sigma$   be a set of substitutions s.t.   $\sigma \in A_{(\SUA)}(\Gamma)$.
	
	Then, from Def.~\ref{d.Combined-Unifier} ({\em\bf Combined Unifier}),   $\sigma \in \sigma_1 \odot \sigma_2$,   where   $\sigma_1 \in A_{\std}(\Gamma_{5.1})$   and   $\sigma_2 \in A_{\acun}(\Gamma_{5.2})$.
	
Suppose there is a term   $t$   in   $\Gamma$   with an alien subterm   $t'$ wrt the theory   $\acun$   (e.g.   $[1,n_a]^{\to}_k \oplus b \oplus c$   with the alien subterm of   $[1,n_a]^{\to}_k$).

Then, from the definition of $\Gamma_2$,   it must have been replaced with a new variable in   $\Gamma_2$.   i.e.,

\begin{equation}\label{e.replace-W-NewVars}	
(\fa t, t')
\left(
\left(
\begin{array}{c}
	(t \in \Gamma) \wedge (t = \_ \oplus \ldots \oplus \_) \wedge \\
	(t' \sqss t) \wedge \msf{ast}(t',t,\acun)
\end{array}
\right)
\Ra
(\ex X)
\left(
\begin{array}{c}
	(X \stackrel{?}{=}_{\acun} t' \in \Gamma_2) \wedge \\
	(X \in \NewVars)
\end{array}
\right)
\right).
\end{equation}

where $\NewVars \subset \Vars \setminus \Vars(\Gamma)$.

Since \txor{} terms do not have free variables from hypothesis, it implies that every free variable in an \txor{} term in $\Gamma_2$ is a new variable:

\begin{equation}\label{e.acun-terms-no-free-vars}
(\fa t, t')
\left(
\left(
	\begin{array}{c}
		(t \in \Gamma_2) \wedge 
		\pure(t,\acun) \wedge \\
		(t' \sqss t) \wedge (t' \in \Vars)
	\end{array}
\right)
	\Ra
	(t' \in \NewVars)
\right).
\end{equation}

Since every alien subterm of every term in   $\Gamma$   has been replaced with a new variable (\ref{e.replace-W-NewVars}), combining it with (\ref{e.acun-terms-no-free-vars}), \txor{}  terms in   $\Gamma_2$   must now have only constants and/or new variables:

\begin{equation}\label{e.newvars-or-constants-Gamma2}
(\fa t, t')
\left(
\left(
	\begin{array}{c}
		\pure(t,\acun) \wedge \\
		(t \in \Gamma_2) \wedge (t' \sqss t)
	\end{array}
\right)	
\Ra
	(t' \in \NewVars \cup \Constants)
	\right).
\end{equation}

Let   $\VarIdP$   be a partition of   $\Vars(\Gamma_2)$   and   $\Gamma_3 = \Gamma_2 \rho$,   such that 

  \[	\Gamma_2 \rho    =    \{  s \stackrel{?}{=}_{\acun} t   \mid   (s \stackrel{?}{=}_{\acun} t := s'\rho \stackrel{?}{=}_{\acun} t'\rho) \wedge s' \stackrel{?}{=}_{\acun} t' \in \Gamma   \}	\]

where   $\rho$   is the set of substitutions where each set of variables  in   $\VarIdP$   has been replaced with one of the variables in the set:

\[
\rho    = 
  \left\{
\begin{array}{cl}
x/X   \mid  
\left(
\begin{array}{c}
	 (\fa Y_1/X_1, Y_2/X_2 \in \rho; \vip \in \VarIdP)
	 		\left( 	 			  
				 	 \begin{array}{c}
					 	 ( X_1, X_2 \in \vip) \Ra \\
					 	 (Y_1 = Y_2) \wedge \\
					 	 (Y_1, Y_2 \in \vip)  
				 	 \end{array}
				 \right)  
\end{array}
\right)
\end{array}
\right\}.
\]

Can there exist a substitution   $X/Y$   in   $\rho$   such that   $Y \in \NewVars$ and   $X \in \Vars(\Gamma)$?

To find out, consider the following two statements:

\begin{itemize}
	
		\item From (\ref{e.replace-W-NewVars}), every new variable   $Y$   in   $\Gamma_2$   belongs to a $\std$-UP in  $\Gamma_2$:
		
		\[	(\fa Y \in \NewVars)( (Y \in \Vars(\Gamma_2) \Ra (\ex t)(\pure(t, \std) \wedge Y \stackrel{?}{=}_{\acun} t \in \Gamma_2)) ).  \]

		\item Further, from hypothesis, we have that \txor{} terms in   $\Gamma$   do not have free variables. Hence, every free variable is a proper subterm\footnote{$t$ is a proper subterm of $t'$ if $t \sqsubset t' \wedge t \neq t'$.} of a purely $\std$ term:

\[
(\fa X \in \Vars(\Gamma))
\left(
			\begin{array}{c}
					(\ex t \in \Gamma) ( (X \sqss t) \wedge 
					\pure(t, \std) \wedge (X \neq t))
			\end{array}
\right).
\]

\end{itemize}

The above two statements are contradictory: It is not possible that a new variable and an existing variable can be replaced with each other, since one belongs to a $\std$-UP, and another is always a proper subterm of a term that belongs to a $\std$-UP. 

Hence,   $\VarIdP$   cannot consist of sets where new variables are replaced by   $\Vars(\Gamma)$.   i.e.,

\begin{equation}\label{e.VarIdP-Prop}
	(\nexists X, Y; \vip \in \VarIdP)
	\left(
		\begin{array}{c}
			(Y,X \in \vip) \wedge (Y \in \NewVars) \wedge \\
			(X \in \Vars(\Gamma)) \wedge (X/Y \in \rho) 
		\end{array}
	\right)
\end{equation}

%For instance, in our example in Section \ref{sss.unification}, if   $\VarIdP = \{ \ldots, \{W,A\}, \ldots  \}$   will not lead to successful unification of   $\Gamma$   since   $W = [1,n_a]^{\to}_{\pk(B)}$   and when it is replaced with   $A$,   it becomes   $A = [1,n_a]^{\to}_{\pk(B)}$.   This implies   $[1, n_a]^{\to}_{\pk(B)} = [1, [1, n_a]^{\to}_{\pk(B)} ]$   which is clearly not unifiable.

Writing (\ref{e.VarIdP-Prop}) in (\ref{e.newvars-or-constants-Gamma2}), we have,

\begin{equation}\label{e.newvars-or-constants-Gamma3}
(\fa t, t')
\left(
\left(
	\begin{array}{c}
		\pure(t,\acun) \wedge \\
		(t \in \Gamma_3) \wedge (t' \sqss t)
	\end{array}
\right)	
\Ra
	(t' \in \NewVars \cup \Constants)
	\right).
\end{equation}

Further, if a variable belongs to a UP of $\Gamma_3$, then the other term of the UP is pure wrt $\std$ theory:
 
\begin{equation}\label{e.NewVars-in-Gamma3-mapto-STDterms}
(\fa X \in \Vars(\Gamma_3), t)
\left(
	\left(
		\begin{array}{c}
			(X \stackrel{?}{=}_{\acun} t \in \Gamma_3)  \vee \\
			(t \stackrel{?}{=}_{\acun} X \in \Gamma_3)
		\end{array}
	\right)
	\Ra
		(X \in \NewVars) \wedge
		\pure(t,\std)
\right).
\end{equation}

Now suppose   $\Gamma_{4.2}    =    \{  s \stackrel{?}{=}_{\acun} t   \mid   (s \stackrel{?}{=}_{\acun} t \in \Gamma_3) \wedge \pure(s,\acun) \wedge \pure(t,\acun)  \}$, 
$\{ V_1, V_2 \}$   a partition of   $\Vars(\Gamma) \cup \NewVars$,   and

\[	\Gamma_{5.2}   =   \Gamma_{4.2} \beta, 	\]

where, $\beta$ is a set of substitutions of new constants to $V_1$: 

 \[	\beta = \{  x/X   \mid   (X \in V_1) \wedge (x \in \Constants \setminus (\Constants(\Gamma) \cup \Constants(\Gamma_{5.1})))  \}.	\]

From hypothesis,   $\Gamma_{5.2}$   is  $\acun$-Unifiable. Hence, we have:

\[	(\fa \sigma)((\fa m' \stackrel{?}{=}_{\acun} t' \in \Gamma_{5.2})(m'\sigma =_{\acun} t'\sigma) \Lra \sigma \in A_{\acun}(\Gamma_{5.2})).  \]

Now consider a   $\sigma$   s.t.   $\sigma \in A_{\acun}(\Gamma_{5.2})$.

From (\ref{e.newvars-or-constants-Gamma3}), we have that \txor{} terms in   $\Gamma_{5.2}$   have only new variables and/or constants and from (\ref{e.NewVars-in-Gamma3-mapto-STDterms}) we have that if $X \in \Vars(\Gamma_{5.2})$, then there exists $t$ s.t. $X \stackrel{?}{=}_{\std} t \in \Gamma_{5.1}$ and $t$ is pure wrt $\std$ theory.
 
Suppose   $V_2 \neq \{ \}$.   Then, there is at least one variable, say   $X \in \Vars(\Gamma_{5.2})$.   This implies that   $X$   is replaced with a constant (say  $x$)   in   $\Gamma_{5.1}$.

Since   $X$   is necessarily a new variable and one term of a $\std$-UP,   this implies that   $x$   must equal some compound term made with $\StdOps$.

However, a compound term made with $\StdOps$ can never equal a constant under the $\std$ theory:

\[	(\not\ex f \in \StdOps; t_1, \ldots, t_n; x \in \Constants)(x =_{\std}   f(t_1,\ldots,t_n)),		\]

a contradiction.

Hence,   $\sigma = \{ \}$,   $V_2 = \{ \}$   and our hypothesis is true that all \txor{} terms in   $\Gamma_{5.2}$   necessarily contain only constants:

\[	(\fa m' \stackrel{?}{=}_{\acun} t' \in \Gamma_{5.2}; x)
			\left(
						\begin{array}{c}
								 (x \sqss m) \vee (x \sqss t) 
								\Ra (x \in \Constants)
						\end{array}
			\right).
\]

\end{proof}

\end{document}